\documentclass[fleqn,twoside]{article}
\usepackage{natbib} 
\usepackage{espcrc2}

\usepackage{psfig}
\usepackage{astro}
\usepackage{graphicx}
\usepackage{natbib} 
\usepackage{journals}     

\newcommand{\AmS}{{\protect\the\textfont2
  A\kern-.1667em\lower.5ex\hbox{M}\kern-.125emS}}

\title{Compact radio cores: from the first black holes to the last}

\author{H. Falcke\address[falcke]{ASTRON, P.O. Box 2, 7990 AA Dwingeloo, The Netherlands, falcke@astron.nl},
 E. K\"ording\address[koerding]{Max-Planck-Institut f\"ur Radioastronomie, Auf dem H\"ugel 69, 53121 Bonn, Germany}, 
N.M. Nagar\address[Nagar]{Kapteyn Institute, Postbus 800, 9700 AV, Groningen, The Netherlands},
}
       
\begin{document}

\begin{abstract}
One of the clearest signs of black hole activity is the presence of a
compact radio core in the nuclei of galaxies. While in the past the
focus had been on the few bright and relativistically beamed sources,
new surveys now show that essentially all black holes produce compact
radio emission that can be used effectively for large radio
surveys. Radio has the advantage of not being affected as much by
obscuration. With the Square Kilometer Array (SKA) these cores can be
used to study the evolution of black holes throughout the universe and
even to detect the very first generation of supermassive black
holes. We start by introducing some of the basic properties of compact
radio cores and how they scale with accretion power. The relative
contribution of jets and radio cores to the Spectral Energy
Distribution is strongest in sub-Eddington black holes but also
present in the most luminous objects. Radio and X-rays are correlated
as a function of black hole mass such that the most massive black
holes are most suited for radio detections. We present a radio core
luminosity function for the present universe down to the least
luminous AGN. The SKA will essentially detect all dormant black holes
in the local universe, such as that in our Milky Way, out to several
tens of Megaparsecs. It will also be able to see black holes in the
making at redshifts $z>10$ for black hole masses larger than
$10^{7}M_\odot$. Finally, we suggest that the first generation of
black holes may have jets that are frustrated in their dense
environment and thus appear as Gigahertz-Peaked-Spectrum (GPS)
sources. Since their intrinsic size and peak frequency are related and
angular size and frequency scale differently with redshift, there is a
unique region in parameter space that should be occupied by emerging
black holes in the epoch of reionization. This can be well probed by
radio-only methods with the SKA.
\vspace{1pc}
\end{abstract}

\maketitle

\section{Introduction}

Clearly, one of the key issues in cosmology centers on the nature of the
very first self-radiating objects in the universe. 
Undoubtedly among these objects will be the first generation of stars
and progenitors of galaxies. However, for some reasons, an almost
unavoidable consequence of galaxy and star formation is the formation
of black holes (e.g., Ferrarese \& Merritt 2000, Gebhardt et
al. 2000). Formation processes can be manifold, through collapse of a
massive star, through collapse of a star cluster (Portegies Zwart et
al.~2004), or through the collapse of inner parts of a
galaxy. Accordingly, the masses spanned by astrophysical black holes
ranges from stellar ($<10^2 M_\odot$) to intermediate ($10^2-10^4
M_\odot$, yet to be confirmed) to supermassive ($ 10^4-10^{10} M_\odot$). As formation
processes of black holes are poorly understood -- we do not even
understand supernovae as progenitors to stellar black holes (Buras et
al. 2003) -- it is very difficult to predict which types of black
holes will have formed first, when, and how.

It is quite possible, however, that the first black holes could have
played an important role in regulating star formation (e.g., Silk \&
Rees 1998) or even contributed to the re-ionization of the
universe. As discussed elsewhere (Furlanetto \& Briggs, this volume)
the so-called epoch of reionization, where the neutral remnant gas of
the big bang is re-ionized by the first radiating objects, is not well
constrained. An early start of reionization as indicated by WMAP
polarization measurements (Kogut et al. 2003) may in fact even require
the contribution from black holes as discussed in Ricotti \& Ostriker
(2003) and Ricotti, Ostriker, \& Gnedin (2003). Also, the suggestion
that $10^{9-10}M_\odot$ black holes may have existed at $z>6$ (e.g., Willott,
McLure, \& Jarvis 2003; Fan et al. 2003) indicates that supermassive
black hole formation was rapid and must have preceded or was coeval
with the epoch of reionization.

The only way to detect black holes at large distances is through their
strong emission, when they light up in the nuclei of galaxies and
appear as quasars or active galactic nuclei (AGN). It is therefore
important to understand how black holes operate and radiate.

The central engine of AGN is usually thought to consist of the black
hole, an accretion inflow, probably accompanied by a hot corona, and a
relativistic jet (Shakura \& Sunyaev 1973, Sunyaev \& Tr\"umper 1979,
Mirabel \& Rodriguez 1999).  The same is true also for black hole
X-ray binaries (XRBs) of stellar masses and it indeed appears that
both mass scales have a lot in common (see Fender, this volume; Fender
\& Belloni 2004).

Standard accretion disk theory has successfully explained the `big
blue bump' in Quasars or the soft X-ray emission of XRBs in the high
state, its emission can usually be described by multi-color black body
radiation from the surface of an optically thick disk.  The jet on the
other hand has an extremely broad spectral energy density (SED). At
least for blazars jet emission can be seen from radio up to
$\gamma$-rays (see e.g., Montigny et al.~1995). The jet SED has two
components, at lower photon energies the emission is thought to be
synchrotron emission, while the X-rays and $\gamma$-rays originate
from inverse Compton processes (see e.g., Gursky \& Schwartz 1977,
Fossati et al.~1998). In X-ray binaries and some BL Lacs synchrotron
might even contribute to the X-rays (Markoff, Falcke, \& Fender 2001)
.

Understanding the SED and its dependence on parameters is crucial for
identifying the right strategy for deep surveys of black holes
throughout the universe, up into and beyond the epoch of reionization.

Both the jet and the disk emission are anisotropic, the jet emission
due to relativistic boosting and the disk due to its geometry and
possibly an obscuring torus.  The orientation dependence was used in
the successful unification schemes of AGN (c.f., Antonucci 1993, Urry
\& Padovani 1995).  Besides the inclination of the symmetry axis of
the system the main parameters governing its appearance are thought to
be the black hole mass and the accretion rate. Additionally the
appearance of the central engine may depend on the spin of the black
hole and the ambient medium.

Using mass and accretion rate AGN and XRBs can be roughly unified into
thermally (high-power, radiatively efficient disks) and non-thermally
(low-power, radiatively inefficient) dominated sources, where the
non-thermal emission may come entirely from the radio jet in quite a
few cases (Falcke, Koerding, \& Markoff 2004).  This makes it
difficult to rely on only one wavelength range to identify AGN
activity.

The radio part of the SED, however, seems to be present in all types
of AGN albeit at different levels. Its great advantage is the low
effect obscuration has on radio observations and less-biased (or
rather well-biased) surveys can be made. 

The Square Kilometer Array (SKA) will be able to observe the faintest
and possibly first AGN up to very high redshifts in the radio. It will
also be able to see almost every supermassive black hole in the nearby
universe and allow us to study the cosmological evolution of the main
AGN parameters.

In this paper we will mainly focus on the information that can be
obtained from the compact radio emission -- the inner jets or ``radio
cores'' of AGN -- making use of the high resolving power of the SKA. First we
discuss some general properties of radio cores and their scaling with
accretion power. We then discuss the local radio-luminosity function
of radio cores and finally point out how one could effectively select
candidates for the very first supermassive black holes in the
universe.

\section{Radio Cores - a Brief Introduction}
So far there is no good evidence that compact radio emission is
produced by accretion flows and where resolved it almost always comes
from relativistic jets.  Jets in AGN are coherent structures with
spatial scales from a few AU up to several megaparsecs. Therefore we
should be able to investigate basic parameters of a jet at different
scales and get similar answers. For example, we can use the large,
extended lobes of radio jets in FR\,II radio galaxies to estimate
their total power, e.g., by calculating their minimum energy content
from synchrotron theory or from their interaction with hot, X-ray
emitting gas and dividing by the life time of the sources (e.g.,
derived from spectral aging). The derived powers (which are often {\em
lower} limits) are very high -- up to $10^{45-47}$ erg/sec (Rawlings
\& Saunders 1991) and thus larger than the total power output of a
typical galaxy.  The only reasonable place where such enormous amounts
of energy can be released is deep in the potential well of a
supermassive black hole.

\begin{figure}
\centerline{
\psfig{figure=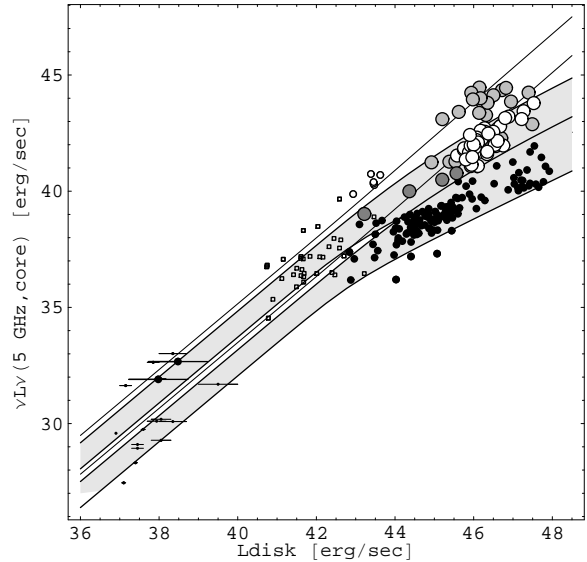,width=0.48\textwidth,bbllx=3.4cm,bblly=17cm,bburx=13.7cm,bbury=27cm,clip=}
}
\vspace{-1cm}
\caption[]{\label{theplot-all}
The correlation between thermal emission from the
accretion disk (with the exception of X-rays this is basically
normalized to their narrow H$\alpha$ emission) and the monochromatic
luminosity of AGN radio cores, spanning the entire observed luminosity
range. Open circles: Radio-loud quasars; small open circles: FR\,I
radio galaxies; open gray circles: Blazars and radio-intermediate
quasars (dark grey); black dots: radio-quiet quasars and Seyferts;
small dots: X-ray binaries; small boxes: detected sources from the
from the ``48 LINERs'' sample (Nagar et al. 2002).  The shaded bands
represent the radio-loud and radio-quiet jet models as discussed in
Falcke \& Biermann (1995)}
\end{figure}

Even though at these larger distances the jet pressure has decreased
enough so that pressure balance between jet and external medium is
reached, it is believed that the implied adiabatic losses are largely
avoided by a constant re-collimation (Sanders 1983) of the jets: the
side-ways lateral expansion leads to a re-collimation shock which in
turn will focus the kinetic energy of the jet back into the forward
direction. Otherwise one would have to account for expansion factors
of $\ga10^6$ between launch and termination of powerful jets. For a
relativistic plasma where adiabatic losses scale as $r^{-2/3}$ this
would translate into an energy loss of four orders of magnitude and
require the jets to start with enormous initial powers of $10^{49-51}$
erg/sec, corresponding to accretion rates of $10^{2-4}M_\odot$/yr and
Eddington luminosities for black holes with a mass of
$10^{10-12}M_\odot$. From all what we know today, this seems to be too
high and one must conclude, that most of the way, the jet does not
suffer adiabatic losses.

The argument based on hotspots still has the problem that it depends
on parameters characterizing the external medium the jet is
interacting with. If we go to low-power jets we will find that most of
them (e.g., in FR\,I radio galaxies, Seyferts) do not even have
hotspots that would provide one with a well-defined, isolated
dissipation region to determine basic jet parameters easily. In the
most extreme cases, however, even powerful jets are stopped inside a
galaxy -- either because they are young or because the ambient density is very
high -- and they appear as very compact radio galaxies of 0.1-10 kpc
in size, named GHz-Peaked-Spectrum (GPS) sources or
Compact-Steep-Spectrum (CSS) sources. The prospects for using such
sources to discover the first black holes is discussed at the end of
this paper.

Every jet, however, should also have an ``inflationary phase'' close
to the nucleus, after leaving the nozzle where the energy density in
powerful jets can be $1-100$ erg/cm$^3$ and above, compared to
$10^{-12}$ erg/cm$^{3}$ in the local ISM.  This is the region where
flat spectrum radio cores are produced and which we will use in the
following to make some quantitative statements. The advantage of radio
cores is that they are largely independent of external conditions and
therefore should be visible in basically all types of sources that
produce relativistic jets. Their independence comes for a price, since
less interaction often means less shocks, less energy dissipation, and
less radio emission. Hence, compact radio cores are much more
difficult to observe than extended lobes and jets, but modern radio
astronomy is now sensitive enough to do just this.

\subsection{Jet-Disk Coupling}
In order to quantify the radio emission we expect from a radio jet
close to the nucleus we will make a few assumptions and resort
to the simple Ansatz made in Falcke \& Biermann (1995), assuming that
every jet is coupled to an accretion disk. A coupled jet-disk system
has to obey the same conservation laws as all other physical systems,
i.e.~at least energy and mass conservation (other conservation laws we
do not use yet). We can express those constraints by specifying that
the total jet power $Q_{\rm jet}$ of the two oppositely directed beams
is a fraction $2q_{\rm j}<1$ of the accretion power $Q_{\rm disk}=\dot
M_{\rm disk}c^2$, the jet mass loss is a fraction $2q_{\rm m}<1$ of
the disk accretion rate $\dot M_{\rm disk}$, and the disk luminosity
is a fraction $q_{\rm l}<1$ of $Q_{\rm disk}$ ($q_{\rm l}=0.05-0.3$
depending on the spin of the black hole).  The dimensionless jet power
$q_{\rm j}$ and mass loss rate $q_{\rm m}$ are coupled by the
relativistic Bernoulli equation (Falcke \& Biermann 1995) for a
jet/disk-system. For a large range in parameter space the total jet
energy is dominated by the kinetic energy such that one has
$\gamma_{\rm j}q_{\rm m}\simeq q_{\rm j}$, in case the jet reaches its
maximum sound speed, $c/\sqrt{3}$, the internal energy becomes of
equal importance and one has $2\gamma_{\rm j}q_{\rm m}\simeq q_{\rm
j}$ ('maximal jet'). The internal energy is assumed to be dominated by
the magnetic field, turbulence, and relativistic particles. We will
constrain the discussion here to the most efficient type of jet where
we have equipartition between the relativistic particles and the
magnetic field and also have equipartition between the internal and
kinetic energy (i.e.~bulk motion) -- one can show that other, less
efficient models would fail to explain the highly efficient radio-loud
quasars but also fail to explain radio cores in low-luminosity AGN.

Knowing the jet energetics, we can describe the longitudinal structure
of the jet by assuming a constant jet velocity (beyond a certain
point) and free expansion according to the maximal sound speed
($c_{\rm s}\la c/\sqrt{3}$). For such a jet, the equations become very
simple. The magnetic field is given by

\begin{equation}
B_{\rm j}=0.3\,G\;Z_{\rm pc}^{-1}\sqrt{q_{\rm
j/l}L_{46}}
\end{equation} 
and the particle number density is
\begin{equation}
n=11\,{\rm cm}^{-3} L_{46} q_{\rm j/l} Z_{\rm pc}^{-2}
\end{equation} 
(in the jet rest frame). Here $Z_{\rm pc}$ is the distance from the
origin in parsec (pc), $L_{\rm 46}$ is the disk luminosity in
$10^{46}$ erg/sec, $2q_{\rm j/l}=2q_{\rm j}/q_{\rm l}=Q_{\rm
jet}/L_{\rm disk}$ is the ratio between jet power (two cones) and disk
luminosity which is of the order 0.1--1 (Falcke, Malkan, \& Biermann
1995) and $\gamma_{\rm j,5}=\gamma_{\rm j}/5$ ($\beta_{\rm
j}\simeq1$). If one calculates the synchrotron spectrum of such a jet,
one obtains locally a self-absorbed spectrum that peaks at

\begin{equation}
\nu_{\rm ssa}=20\,{\rm GHz}\;{\cal D}{\left(q_{\rm j/l}L_{46}\right)^{2/3}
\over Z_{\rm pc}}\,\left({\gamma_{\rm e,100} 
\over\gamma_{\rm j,5} \sin i}\right)^{1/3}.
\end{equation}
Integration over the whole jet yields a flat spectrum with a
monochromatic luminosity of

\begin{eqnarray}\label{radioopt}
L_{\nu}&=&{ 1.3\cdot 10^{33}}\,{{\rm erg}\over
{\rm s\, Hz}}\;\left({q_{\rm j/l} L_{46} }\right)^{17/12}
{\cal D}^{13/6}\sin i^{1/6} \nonumber\\
&&\times\gamma_{\rm
e,100}^{5/6} \gamma_{\rm j,5}^{11/6},
\end{eqnarray}
where $\gamma_{\rm e,100}$ is the minimum {\it electron} Lorentz
factor divided by 100, and ${\cal D}$ is the {\it bulk} jet Doppler
factor. At a redshift of 0.5 this luminosity corresponds to an
un-boosted flux of $\sim100$ mJy. The brightness temperature of the jet
is

\begin{eqnarray}
{T}_{\rm b}&=&1.2\cdot 10^{11}\, {\rm K}\; {\cal D}^{4/5}{\left({
{\gamma_{\rm e,100}}^2 q_{\rm j/l} L_{46} \over
\gamma_{\rm j,5}^2 \beta_{\rm j}}\right)^{1/12}}\nonumber\\
&&{\times\sin i^{5/6}}
\end{eqnarray} 
which is almost independent of all parameters except the Doppler
factor. A more detailed calculation of this can be found in Falcke \&
Biermann (1995).

A basic conclusion is that compact radio cores should exist
essentially in all AGN, subject mainly to a range in accretion
powers. Hence, we can expect many faint radio cores in nearby
low-luminosity AGN (LLAGN) and a few bright radio cores in the most
distant and brightest AGN that overall share almost the same
properties: compact with a flat radio spectrum. This has been
observationally verified by radio surveys of different types of black
holes, starting at stellar masses and low powers (see Fender, this
volume), via LLAGN (next section), all the way up to the brightest
quasars. Figure~\ref{theplot-all} shows radio core fluxes as a
function of optical/X-ray luminosity, illustrating the enormous range
compact radio cores occupy.

\subsection{Power Unification}
Figure~\ref{theplot-all} seems to indicate that the radio luminosity
of black holes simply scales with the thermal component, i.e. the
accretion disk luminosity. However, the actual picture is more
complicated as accretion flows and their related emission may not show
a simple linear scaling. Recent results suggest that accreting black
holes can exist in mainly two distinct states. Which state it is in
depends mainly on the accretion rate. In the low-power state the
standard optically thick disk is probably truncated and a radiatively
inefficient optically-thin accretion flow exists in the inner regions
of the flow (see e.g., Esin et al.~1997, Poutanen~1998). In the high
state the optically-thick geometrically thin accretion flow seems to
continue to the innermost stable orbit. For XRBs the transition
appears to happen at an accretion rate of roughly 10\% Eddington (see
e.g., Narayan \& Yi 1995). Recent work suggests that this transition
can already happen at 2\% of the Eddington rate (Maccarone 2004), and
that there is a hysteresis in the critical accretion rate depending on
which direction the transition is going along (Maccarone \& Coppi
2003, Fender \& Belloni 2004). For AGN the position of the transition
is also compatible with a few percent Eddington (Ghisellini \& Celotti
2001).

At least the low power state is normally connected with relativistic
outflows. As mentioned above the jet will contribute significantly to
the SED for XRBs and AGN (Falcke \& Markoff 2000, Markoff et al.~2001,
Fender~2001, Yuan, Markoff, Falcke 2002). As the accretion flow is
radiatively inefficient in the innermost regions the jet can even
dominate the overall emission (see e.g., Fender, Gallo, Jonker 2003,
Falcke et al.~2004).  This concept of jet-domination can be used to
unify stellar and super-massive black holes.

\begin{figure}
\begin{center}
\resizebox{0.49\textwidth}{!}{\includegraphics{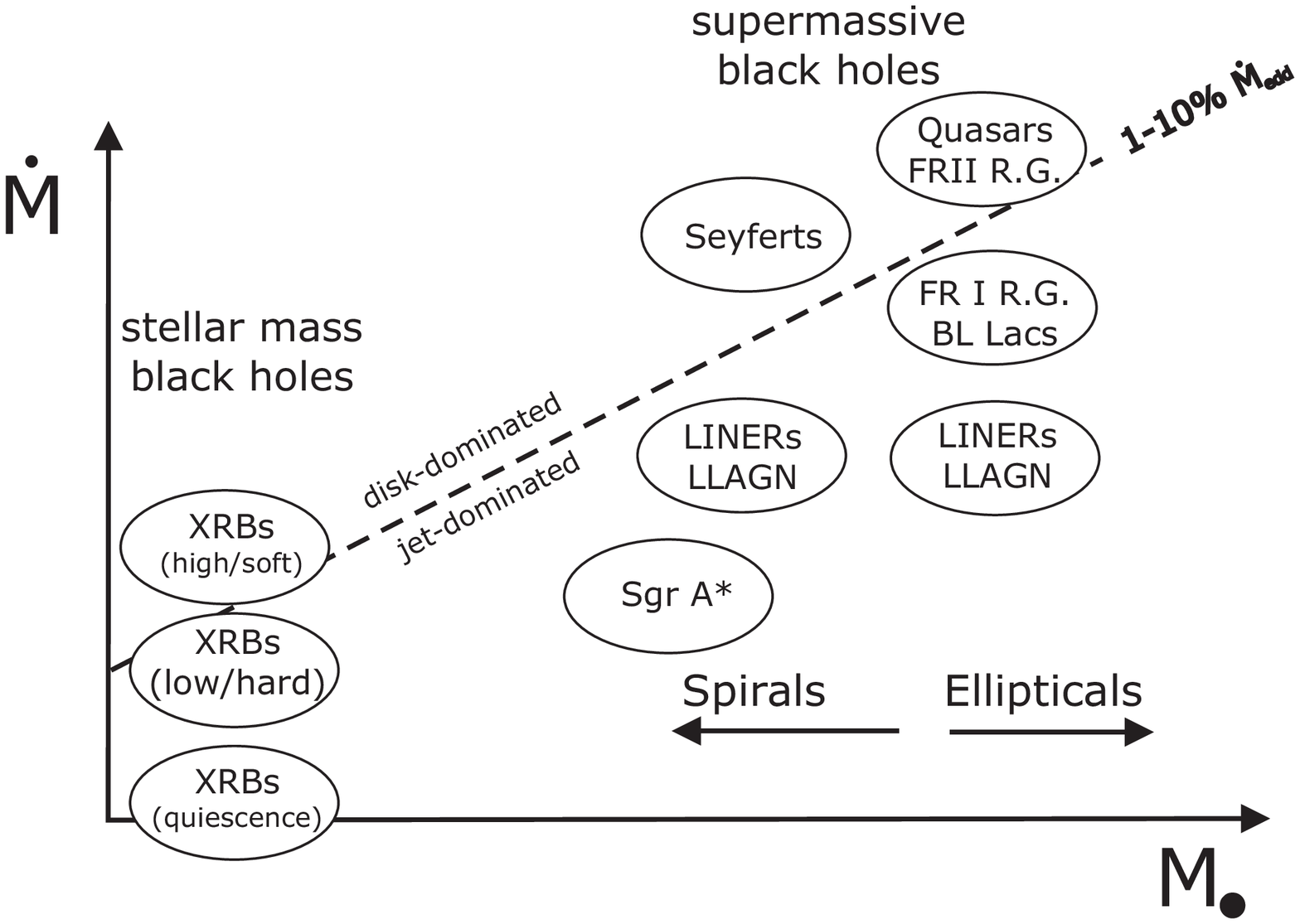}}
\end{center}
\vspace{-1cm}
\caption[]{
A proposed unification scheme for accreting black holes in the mass
and accretion rate plane. Above a few percent of the Eddington
accretion rate, the systems are proposed to be dominated by disk
emission, while below they are inherently dominated by jet emission
(RG=radio galaxy). This unification scheme extends the standard
inclination-based unified schemes
(Antonucci 1993, Urry Padovani 1995).}\label{scheme}
\end{figure}

The main assumptions used for the unification scheme are:
\begin{enumerate}
\item The jet and the accretion flow is a coupled symbiotic system. If one of them exists the other will also be present (e.g., Falcke \& Biermann 1995). 
\item Below a critical accretion rate $\dot{M}_{\mathrm{crit}}$ which is a small fraction of the Eddington accretion rate, the innermost parts of the accretion flow become radiatively inefficient.
\item For $\dot{M} \leq \dot{M}_{\mathrm{crit}}$ the overall emission will be dominated by the jet. 
\end{enumerate}

This requires a better classification of AGN into disk or jet
dominated classes and a single basic SED cannot be assumed.  As
mentioned above black hole XRBs appear mainly in two distinct states:
the low/hard state characterized by a hard power-law (e.g., Nowak
1995) and the high state where the SED is dominated by a thermal
component. The hard power-law of the low-hard state is often explained
by Comptonization in a hot corona or the optically thin accretion flow
(see, e.g.~Shapiro, Lightman, \& Eardley 1976, Sunyaev \& Tr\"umper
1979) but can also be explained as synchrotron emission from a
relativistic jet (Markoff et al.~2001). The latter interpretation is
further supported by the finding of a fairly universal radio/X-ray
correlation in the low/hard state (Gallo, Fender, \& Pooley 2003)
which can be well explained with a jet model (e.g., Markoff et
al. 2003, Falcke et al. 2004). Thus, we identify the low/hard state as
jet-dominated while the soft state is assumed to be disk-dominated.
As there are many different classes of AGN the classification in
disk/jet dominated sources is more complicated than for the stellar
counterparts. The current classification is summarized in
Fig.~\ref{scheme}. For a more detailed analysis see Falcke et
al.~(2004).

\subsection{The Radio/X-ray correlation}
Clearly, if the SED is dominated by one component -- suggested for the
jet-dominated low-power system, different wavelengths should be
related in a defined manner, subject to scaling with mass and accretion
rate. This can be used to predict which wavelength range is best for
detecting certain types of AGN; vice versa one could also say that
every wavelength range somehow pre-selects certain ranges in mass and
accretion rate.

For example, it has been found recently that radio and X-rays from
black holes are correlated in a very specific, mass-dependent manner
(Falcke et al.~2004; see also Merloni, Heinz, \& di~Matteo 2003). This
can be easily understood using the jet domination idea, where the
radio is from the optically thick part of the spectrum, while X-rays
come from the optically thin part (other explanations have also been
put forward, however). Using scaling laws for the jet SED it can then
be shown that the X-ray luminosity is expected to depend on the radio
luminosity and the black hole mass, $M_\bullet$, as:
\begin{equation}
L_{\rm X} \propto L_{\rm R}^{1.38} M_\bullet^{0.81}.
\label{eqRXScale}
\end{equation} 

\begin{figure}[t]
\begin{center}
\resizebox{0.48\textwidth}{!}{\includegraphics{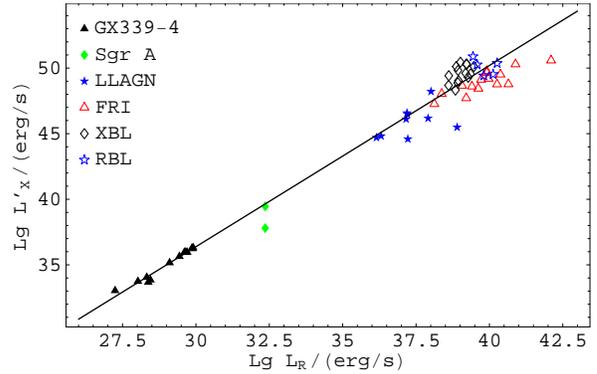}}
\end{center}
\vspace{-1cm}
\caption[]{
Radio/equivalent X-ray correlation for a sample of jet dominated AGN and XRBs. The X-ray flux has been adjusted to correspond to a black hole mass of 6$M_{\odot}$.  
}\label{xprimer-correlation}
\end{figure}

Of course, for many AGN the X-ray emission probably does not originate
from synchrotron processes, still the correlation seems to be
applicable for a large range of sources.  If one corrects for the mass
dependence of the X-ray/radio correlation according to
Eq.~\ref{eqRXScale} one finds a universal radio/X-ray correlation for
all jet dominated sources from XRBs up to AGN as shown in
Fig.~\ref{xprimer-correlation}. Put differently: optically thick
(radio) emission, optically thin (X-rays or optical) emission, and
black hole mass form a fundamental plane for low-power black holes.

\subsection{Observational Consequences}\label{consequences}
\begin{figure*}
\centerline{\psfig{figure=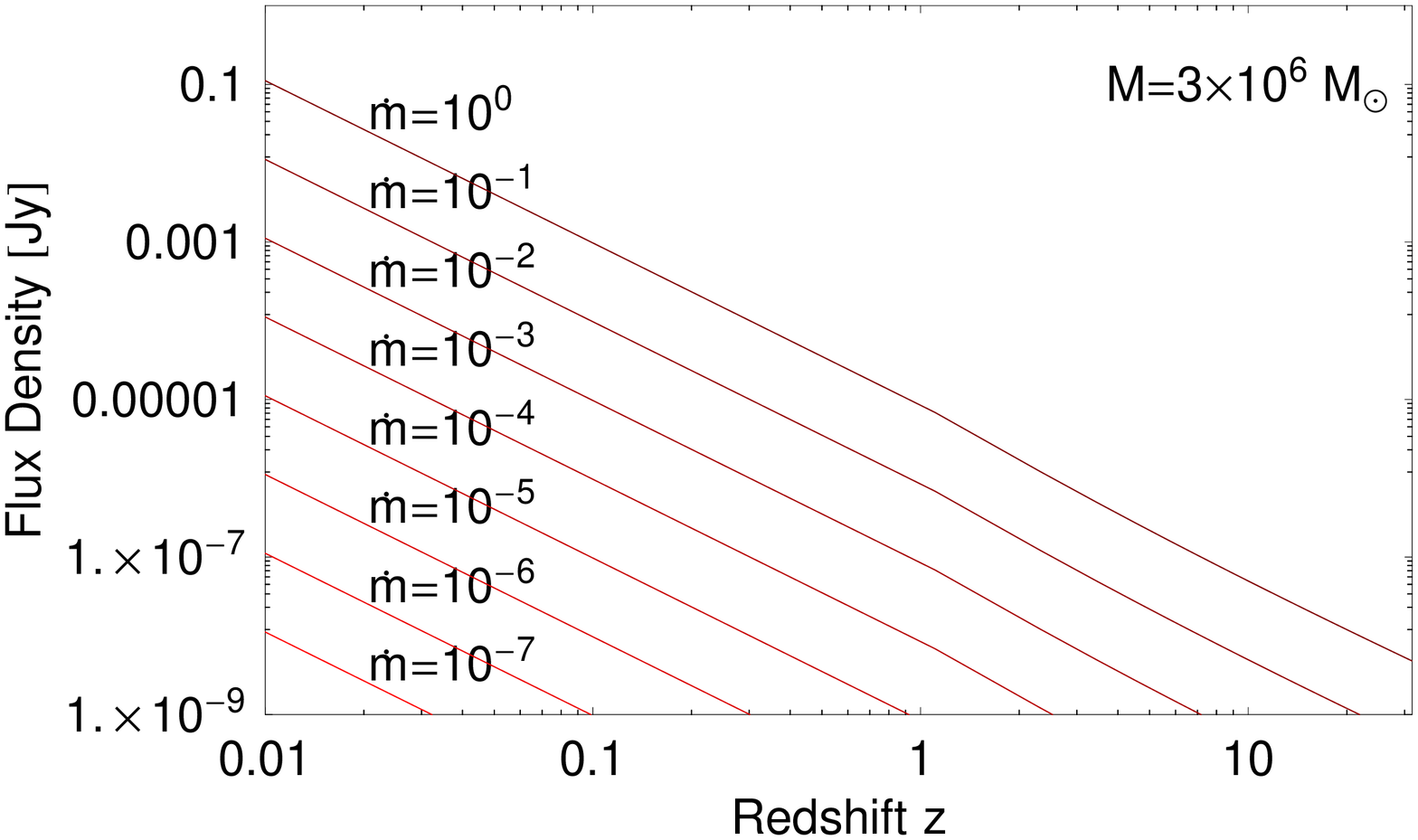,width=0.49\textwidth}\psfig{figure=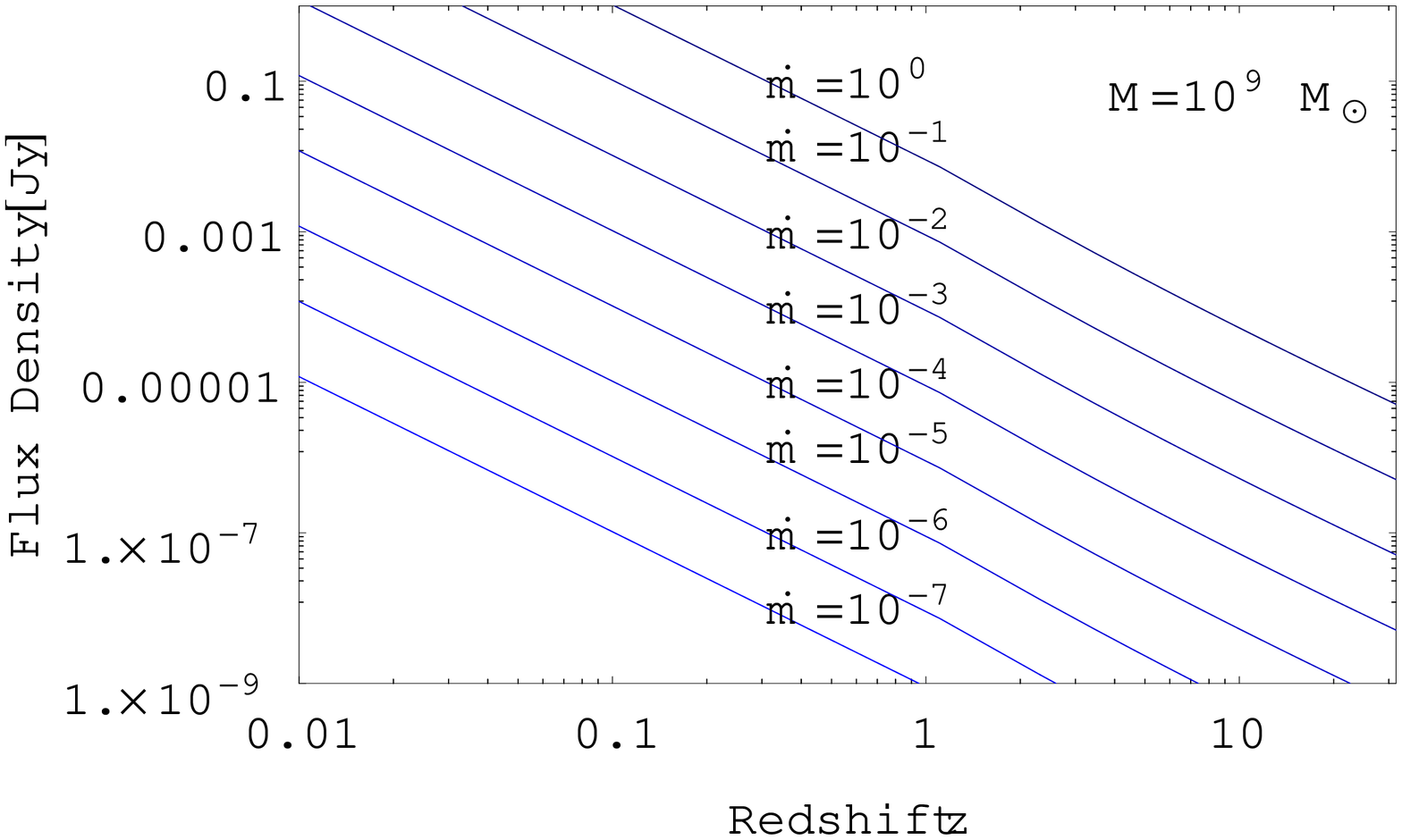,width=0.49\textwidth}}
\vspace{-1cm}
\caption{\label{fluxvsz} 
Predicted fluxes of radio cores in AGN observed at 5 GHz as a function
of redshift for a range of Eddington accretion rates (solid lines from
top to bottom) and two black hole masses: $3\times10^6 M_\odot$ (left)
and $10^9 M_\odot$ (right). An inclination angle of $50^\circ$ is
assumed, i.e. negligible beaming.}
\end{figure*}
From the radio/X-ray correlation we can already deduce which compact
sources we can expect to observe with the SKA at radio frequencies.
\begin{itemize} 
\item XRBs: Compared with AGN all XRB are X-ray loud
(up to a factor $10^8$ more X-ray flux for a given jet power), but
relatively weak in the radio regime hence X-ray telescopes are better
suited for low-mass black holes. However, as the radio flux can be
relativistically beamed in some sources the SKA will allow us to
observe these sources not only in our Galaxy, but also in nearby
galaxies.
\item AGN: With their large total jet power and high black hole masses 
these sources can be radio-loud. Additional relativistic beaming
allows the detection of AGN up to very high redshifts.  The SKA will
boost AGN research, as it will allow to observe AGN of lower
luminosities and at larger distances.  
\item Intermediate mass black
holes (IMBHs): If this intriguing class of objects should exist it
would lie in between AGN and XRBs on the radio/X-ray correlation. With
new generation radio telescopes these objects should be visible in
nearby galaxies. As Maccarone (2004) points out, IMBHs in the center
of globular clusters are easier detected in the radio
regime.  Their X-ray luminosity would be further reduced as low 
accretion rate systems might not start accelerating relativistic 
particles (e.g. Sgr A$^*$ in its quiescent state).
\end{itemize} 

In Fig.~\ref{fluxvsz} we show predicted radio core fluxes as a
function of redshift for a range of radio core models at 5 GHz and for
different masses and accretion rates relative to the Eddington
rate. The fluxes are predicted on the basis of Eq.~\ref{radioopt} for
an inclination angle of $i=50^\circ$ and taking the mass scaling into
account as discussed in Falcke et al.~(2004). The left figure
corresponds to a black hole similar to Sgr A* in the Milky Way (see,
e.g., Melia \& Falcke 2001) and the right one to a black hole in a
massive elliptical galaxy. The bottom line on the left corresponds to
the activity level in Sgr A*, while the top one on the right
corresponds to a bright radio-loud quasar. Assuming that the SKA,
after long integration, can reach levels as low as 10 nJy, we see that
Sgr A*-like objects, i.e., essentially every supermassive black hole
in the local universe, can be detected out to 40 Mpc.

Quasar radio cores, on the other hand, can essentially be seen all the
way out to redshifts of $z>30$. For a survey with detection limits
around 0.1 $\mu$Jy we can even see black holes in normal galaxies
shining at the Eddington rate out to $z\simeq8$. Assuming that the
formation of galaxies proceeds rapidly in the early phases of the
universe at $z>10$ and black holes accrete at the Eddington limit, we
could in principle see the associated radio cores as soon as a mass of
$10^7M_\odot$ is reached. In section \ref{firstbh} we will, however,
discuss another possibility to detect black holes in the making. Note
that the predictions made here only assumed an average inclination
angle. Obviously, some of these sources will be beamed and hence a few
lower mass black holes could be seen as well as flat-spectrum
sources. A crucial prerequisite is, of course, the presence of
sensitive long baselines in the SKA, to isolate the high-brightness
temperature cores from diffuse star formation.

\section{SKA and the Weak AGN Tail}
\label{secrlf}

\begin{figure}[ht]
\resizebox{\columnwidth}{!}{
   \includegraphics{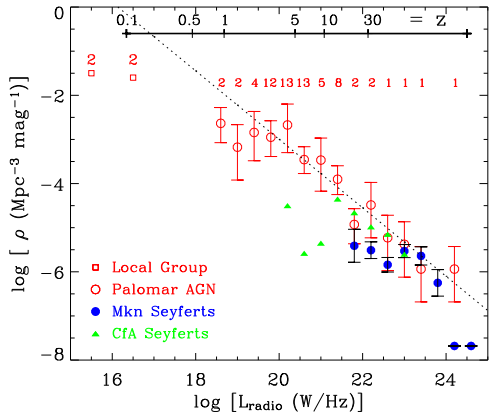}
}
\caption{The 15~GHz nuclear radio luminosity function (RLF) of the 150~mas-scale
radio nuclei in the Palomar sample (red open circles, with the number
of galaxies in each bin listed above the symbol) as compared to the
RLFs for Markarian Seyferts (RLF from Meurs \& Wilson 1984) and CfA
Seyferts (RLF calculated by Ulvestad \& Ho (2001) using data from
Kukula et al.~1995).
The dashed line is a power-law ($-0.78$) fit to the Palomar RLF (excluding
the two lowest radio luminosity points).
Also shown is the estimated nuclear RLF of galaxies in the
local group (red squares, with 2 galaxies in each of the two bins; see text).
The upperscale illustrates the detection limits of the SKA 
(using an r.m.s noise of 0.3$\mu$Jy in 1~hr of imaging, and a 3$\sigma$ 
 detection threshold) of similar AGNs at redshifts of 0.1 to 30.}
\label{figrlf}
\end{figure}

Some idea of the high-z AGN population accessable to SKA can be
obtained from extrapolating deep radio surveys of local AGNs.
Currently the deepest radio search for AGNs in a large well-defined
sample of nearby galaxies is a VLA and VLBA survey of the Palomar
sample of all $\sim$480 nearby bright (B$_{\rm T}\,<$12.5 mag)
northern galaxies. Optical spectroscopy of the sample (Ho, Filippenko,
\& Sargent 2003) shows that it is made up of roughly 197 AGNs (almost
all with L$_{\rm H\alpha}~\leq$ 10$^{40}$ erg s$^{-1}$,
i.e. low-luminosity AGNs or LLAGNs) 206 nuclei with H~II type nuclear
spectra, and 53 absorption line nuclei.  Effectively all galaxies have
been surveyed to detection limits of 1~mJy with the VLA (15~GHz;
0{\farcs}15 resolution), with follow-up VLBA observations (5~GHz,
2~mas resolution) confirming the AGN nature of all VLA detections with
flux $>$2.7~mJy (Falcke et al. 2000, Nagar et al.~2004).

Most of the $\sim$70 VLA-detected radio nuclei are compact or at least the
AGN-related radio emission is primarily from within the central arcsec.
The radio detected nuclei have been used to derive a
\textit{nuclear, i.e. AGN core} 15~GHz radio luminosity function (RLF) which is
plotted in Fig.~\ref{figrlf} as open red circles{\footnote{The RLF
has been computed via the bivariate optical-radio luminosity function
(following the method of Meurs \& Wilson 1984), after correcting for
the incompleteness (Sandage, Tammann, \& Yahil 1979) of the RSA catalog (from which the
Palomar sample was drawn). Errors were computed following the method
of Condon (1989).}}; see also Ulvestad \& Ho (2001, RLF for Palomar Seyferts), and
Filho (2003, $PhD$ thesis; RLF for Palomar sample using diverse radio surveys).
At the highest luminosities the RLF is in good agreement with
that of `classical' Seyferts (Fig.~\ref{figrlf}). We use the 
other RLFs without correction for frequency or nuclear vs. total
AGN emission since our RLF is derived from nuclei which show 
relatively flat radio spectra from 1.4--15~GHz and which are
relatively compact (i.e. sub-arcsec).
At lower luminosities, the sample extends the RLF of AGNs by more than 
three orders of magnitude.
A fit to the Palomar RLF (excluding the two lowest luminosity
bins; see below) yields:
\begin{eqnarray}
{\log}\rho &=&\left(12.5 - 0.78\times\,{\rm log}\,\left({\rm L}_{\rm radio}\,[{\rm W\,Hz}^{-1}]\right)\right)\nonumber\\ &&[{\rm Mpc}^{-3}\,{\rm mag}^{-1}]
\end{eqnarray}
An approximate RLF for the nuclei of the local group of galaxies is also plotted
in Fig.~\ref{figrlf}.
At $\sim10^{15}$ W Hz$^{-1}$ one has two galaxies: the Milky Way and M~31.
At $\sim10^{16}$ W Hz$^{-1}$, NGC~205 and NGC~598 (see Nagar et al, 2004).
These lowest points, and our extrapolations of the radio non-detections,
support a lower power turnover for the RLF, though the apparent turnover could 
also be due to the incompleteness of the radio survey.

\begin{figure}[ht]
\resizebox{\columnwidth}{!}{
   \includegraphics{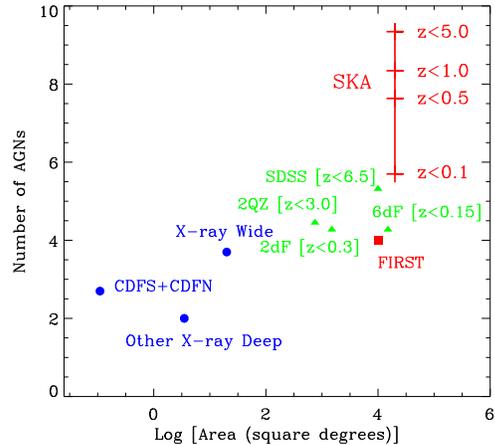}
}
\caption{A comparison of the number of AGNs expected to be identified by several current large area 
or deep surveys as a function of their sky coverage. The typical redshift limit probed by the
survey is also marked in some cases.
X-ray surveys in blue: CDFS and CDFN are the Chandra Deep Field North and South. ``X-ray wide'' indiactes a shallow X-rays survey with a much wider field of view.
Optical surveys in green: these consist of the Two-degree-field galaxy redshift survey (2dF),
the Two-degree-field QSO survey (2QZ), the Six-degree-field survey (6dF) and the Sloan
Digital Sky Survey (SDSS). 
Radio surveys in red; the predicted number counts for a VLBI-SKA `one-hemisphere' survey 
(1~hr per pointing) out to different redshifts are indicated.}
\label{figskapred}
\end{figure}

The upper bar in Fig~\ref{figrlf} illustrates the detection limits of
the SKA (using an r.m.s noise of 0.3$\mu$Jy in 1~hr of imaging, and a
3$\sigma$ detection threshold) of a similar sample of AGNs at
redshifts of 0.1 to 30.  The most luminous sources here would be
detectable out to $z\,=\,30$, while at $z\,=\,0.1$ radio nuclei in
AGNs can be detected down to levels of 10 times Sgr~A$^*$.  We have
used these SKA detection limits, in combination with our RLF and the
SDSS galaxy luminosity function (determined for z$\,=\,0-3.5$) to
estimate the number of AGNs identified by the SKA (assuming a VLBI SKA, so
as to easily distinguish AGN from starbursts) and compared these with
several large optical and hard X-ray surveys
(Fig.~\ref{figskapred}). The AGN counts for these other surveys were
estimated from their total galaxy counts in conjunction with intial
results on the incidence of AGN in those surveys.  Clearly, an SKA
survey will find an unprecedented number of AGNs over all redshift
ranges.

\begin{figure}[ht]
\resizebox{\columnwidth}{!}{
   \includegraphics{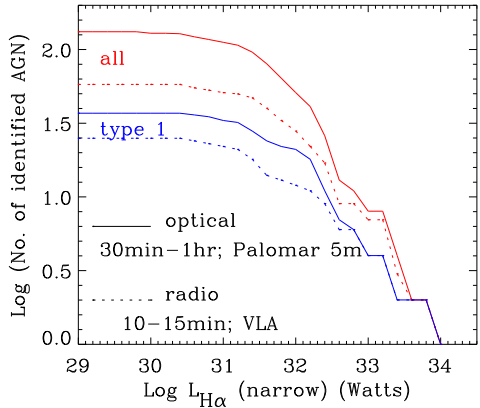}
}
\caption{The cumulative number of `definite' (i.e. those detected in the hard X-ray)
AGNs identified by optical spectroscopy (solid lines; $\sim$4~hr on the Palomar 5~meter) 
and radio imaging (dashed lines; $\sim$20~min at the VLA), 
as a function of the luminosity of the narrow H$\alpha$ line. 
The blue lines are for only type 1 nuclei (i.e. nuclei with broad H$\alpha$
emission) and the red lines for both type 1 and type 2 nuclei.}
\label{figoptvsrad}
\end{figure}

The effectiveness of the SKA in detecting AGN is illustrated above and in the
previous section. However, it is also valid to ask how complete such an
AGN sample would be, both intrinsically, and as compared to X-ray or
optically identified AGN.
To illustrate the effectiveness of radio surveys for AGNs at the lowest luminosities,
we plot (Fig.~\ref{figoptvsrad}) the cumulative number of definite AGNs (here we 
take the presence of a hard X-ray source as a definite sign of AGN activity)
identified by optical spectroscopy (typically 4~hr on a 5~meter telescope) versus
radio imaging (typically 20~min at the VLA). Clearly, the radio survey picks up
a highly significant fraction of definite AGNs, and in fact, in the absence of
hard X-ray data is a more reliable indicator of true AGN activity than optical
spectroscopy, especially at these low luminosities.

\section{The First Black Holes}\label{firstbh}
\subsection{HI spectroscopy}

One of the big goals of the SKA, however, is the identification of the
very first and hence youngest black holes in the universe. This
requires detecting them in the redshift range 7--20. Conventional
optical spectroscopy becomes difficult in this regime due to the
pronounced Ly$\alpha$ troughs (e.g., Becker et al. 2001). Therefore,
the prospects of radio spectroscopy, using the 21 cm hydrogen line, at
very high-redshifts has become an important topic for the next
generation of radio telescopes. Carilli, Gnedin, \& Owen (2002)
discuss the detectability of HI absorption towards radio galaxies at
redshifts $z=8-10$. The conclusion is that depending on the quasar
evolution model hundreds to thousands of bright QSOs should be
detectable at these redshifts in the radio, assuming radio powers
comparable to Cygnus A and black hole masses as high as several $10^9
M_\odot$.


Of course, even with the SKA detecting these early black holes is not
an easy task, requiring a lot of integration time. Given that the SKA
will be able to see essentially every radio galaxy that has existed
within the accessible universe, the number of candidate sources is
enormous and effective survey and selection criteria have to be
developed. After all, one of the big advantages (and curses) of SKA
will be its powerful surveying capability due to its large field of
view.

Consequently, we have to discuss how the first black holes will look
like and what would distinguish them from other radio sources. While
such an attempt will always be speculative it is not futile. In
contrast to the formation of the first stars, the energy release from
black holes is primarily release of gravitational energy and not due
to nuclear synthesis. Unknown metallicity effects should not play as
much a role as it will for the first stars. This is even more true for
the radio emission which comes from a hot, fully ionized, and
relativistic plasma. We can thus expect that the appearance of the
first black holes themselves will not be very much different from what
we see today. Modifications in the appearance are mainly expected from
cosmological effects (extreme redshift), a denser ambient photon field
(inverse Compton cooling), and a different -- presumably much denser
-- environment in the emerging nuclei of forming galaxies.

In section \ref{consequences} we discussed how the inner parts of the
jet, the radio core, can be detected. In the following we will
concentrate more on the extended emission, even though this could be
confined to compact scales as well, as we argue below.

\subsection{High-redshift radio galaxies}
\begin{figure}
\centerline{\psfig{figure=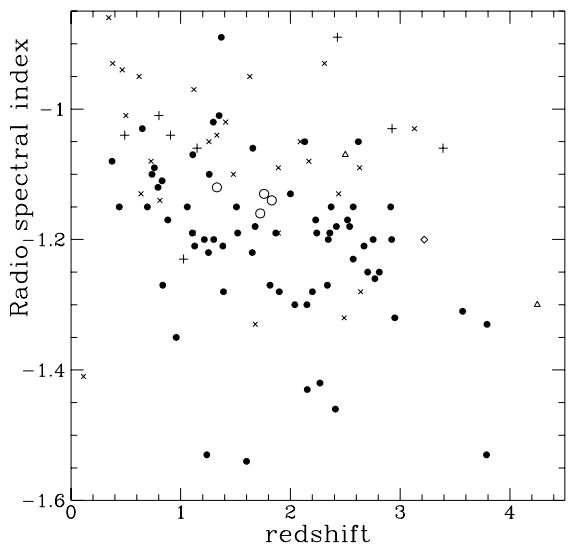,width=0.49\textwidth}}
\vspace{-1cm}
\caption{\label{rottgering97} Radio spectral index plotted against redshift
for 108 radio galaxies. Filled dots are from the ultra-steep spectrum
(USS) sample of R\"ottgering et al. (1997). The spectral indices are
determined between the lowest available frequency above 150 MHz and 5
GHz. The open circles are also from the R\"ottgering et al. (1997) USS
sample with redshifts obtained at Lick Observatory by McCarthy \& van
Breugel (1994, private communication).  The crosses ($\times$) are
from the Molonglo sample of McCarthy (1990a,1990b), the pluses ($+$)
are Bologna sources (McCarthy 1991, Lilly 1988), the triangles are
from the 8C survey (Lacy et al. 1994, Lacy 1992) and the diamond is
6C\,1232+39 (Eales \& Rawlings 1993).  Figure reproduced from
R\"ottgering et al. (1997) -- Astronomy \& Astrophysics}
\end{figure}

One fairly successful technique to find high redshift radio galaxies
has been to select on steep spectral indices. The radio emission from
typical radio galaxies is normally dominated by optically thin
synchrotron emission from their lobes. The spectral index in the radio
then reflects the particle energy distribution; both are a power law
over a wide range of energies, respectively frequencies.  Beyond a
certain break energy and frequency the power law steepens due to
cooling losses. Clearly, with higher redshifts this steepening will
occur at lower and lower observing frequencies. In addition the higher
photon density in earlier epochs may lead to enhanced cooling and
lower break frequencies such that high-z radio galaxies appear to have
steeper spectra. A similar effect is expected if the radio source
spectra have a curved intrinsic shape as discussed, for example, by
Blundell, Rawlings, \& Willott (1999).

Various surveys (see, e.g., R\"ottgering 1997) indicate that indeed
spectral index and redshift are correlated
(Fig.~\ref{rottgering97}). Standard selection procedures are to
compare radio surveys at two or more frequencies, select the ones
with steeper spectral index ($\alpha<-1$, where the flux density as a
function of frequency is $S_\nu\propto\nu^\alpha$), determine their
structure and position with arcsecond-resolution radio observations,
and obtain their spectral index with optical spectroscopy. The latter
part is a tedious process and may become prohibitive for the large
number of sources the SKA is going to provide. Moreover, it is not
clear whether the trend of steeper spectral index with higher redshift
continues, especially as the spectral index also depends on size and
power of a radio galaxy (e.g., Dennett-Thorpe et al. 1999). Still,
with its large spectral and instantaneous bandwidth coverage the SKA
will deliver a huge number of ultra-steep spectrum sources and very
interesting high-z candidates (see chapter by Jarvis \& Rawlings, this volume).

\subsection{The Very First Black Holes}
However, one can ask whether indeed all these arguments apply to the
very first generation of black holes -- those where galaxy and central
black hole are in the process of being formed. The growth process of
black holes can be rather rapid -- within 300-600 Million years a
black hole can grow to more than $10^9 M_\odot$ if it accretes at the
Eddington rate. As argued in section \ref{consequences} this means
even at $z>10$ the compact radio emission from inner jets can be seen;
however, what happens to the extended emission?

Given that the densities of the ISM in the host galaxy must be
extremely high, it is not clear that the jets will be able to even
penetrate into the -- also very dense -- intergalactic medium. If so,
the jet will develop very strong and very bright hot spots {\em
inside} the forming galaxy similar to what we see in GPS sources.

GPS sources show a highly peaked spectrum in the GHz regime, where the
deficiency at low frequencies is most likely due to synchrotron
self-absorption, only in a few cases free-free absorption also
contributes. Hotspots, if understood as homogenous synchrotron
sources, should follow some simple relations. Especially, one expects
that the turn-over (or the self-absorption) frequency decreases with
increasing size. This can be seen in studies of samples and even in
individual sources -- very impressively for example in the evolution
of the radio hotspots of III Zw 2, peaking at 43 GHz (Brunthaler et
al. 2003).

\begin{figure}
\centerline{\psfig{figure=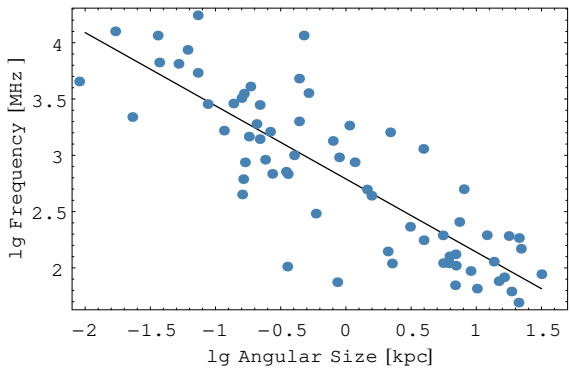,width=0.49\textwidth}}
\vspace{-1cm}
\caption{\label{odeaplot} 
The intrinsic turnover frequency vs. linear size of radio sources in the Fanti et
al. (1990) CSS sample and the Stanghellini et al. (1996) GPS
sample. Adapted from O'Dea \& Baum (1997) and O'Dea (1998).}
\end{figure}
\begin{figure*}
\centerline{\psfig{figure=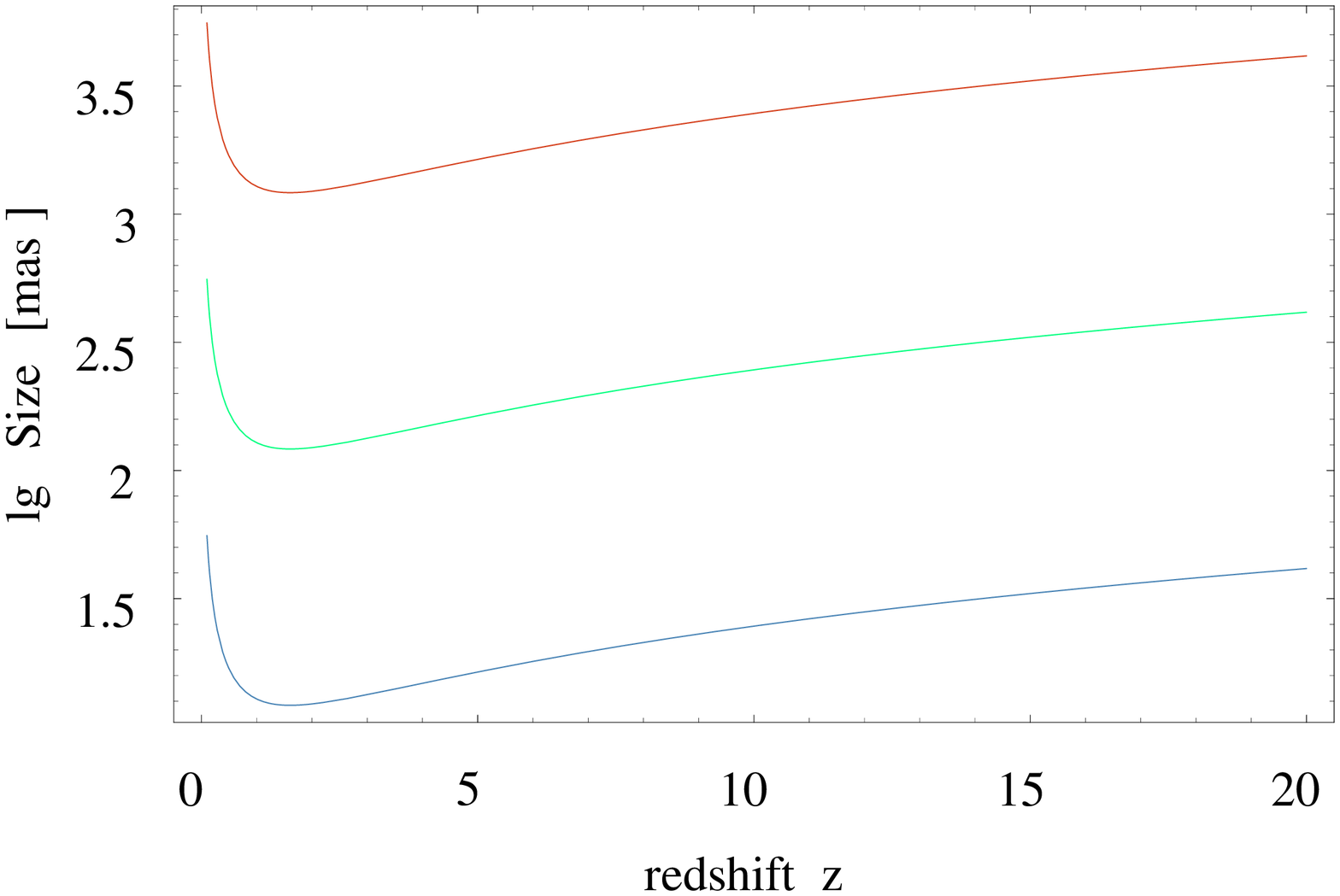,width=0.49\textwidth}\psfig{figure=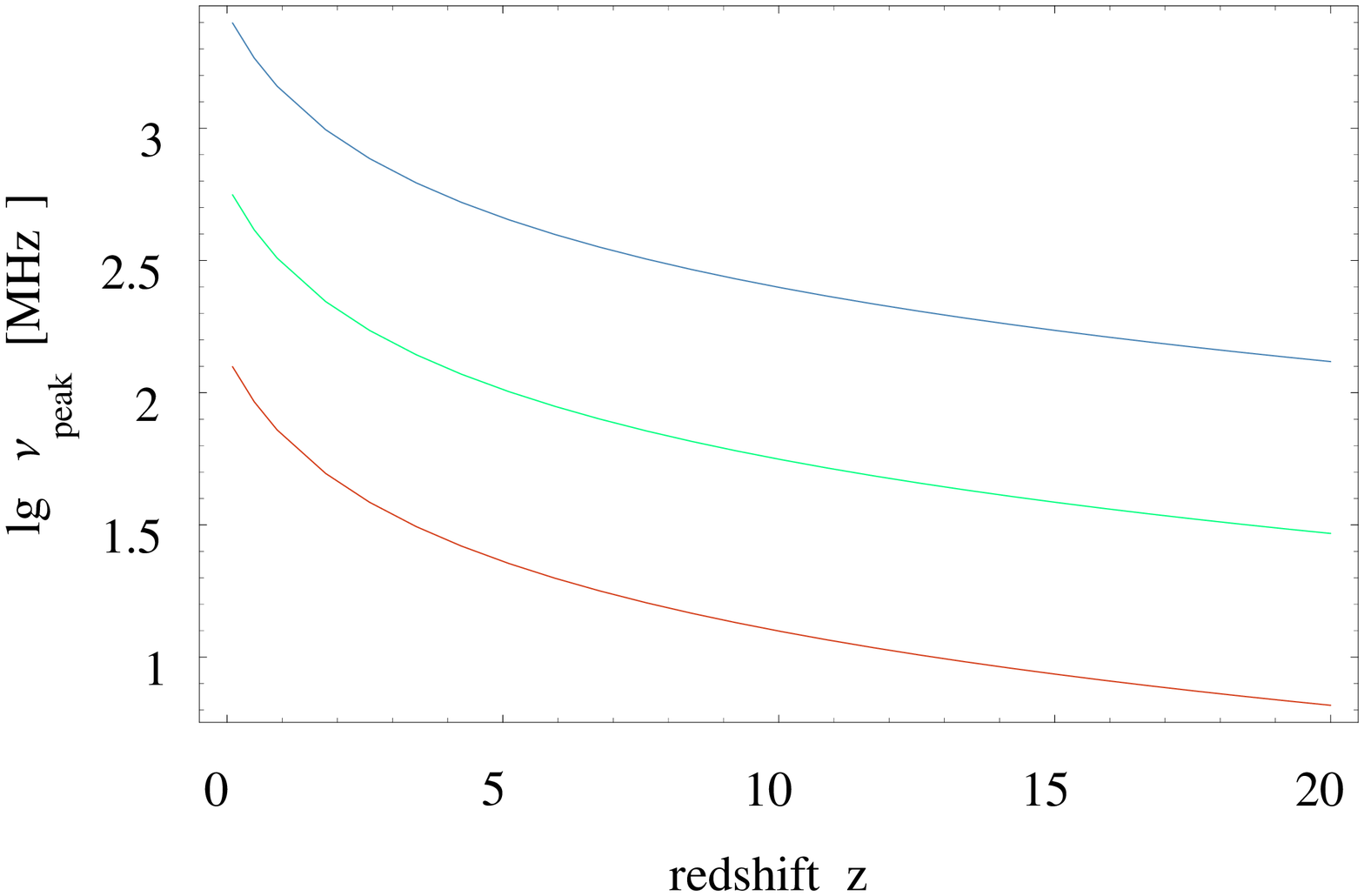,width=0.49\textwidth}}
\vspace{-1cm}
\caption{\label{sizevsz} 
Within a standard cosmology ($\Omega_{\rm m}=0.3$,
$\Omega_{\lambda}=0.7$, $H_0=72$ km/sec/Mpc) a fixed angular size
(left) and peak frequency (right) of a source scale differently as a
function of redshift. The figure shows three examples for sources with
0.1 (blue), 1 (green), and 10 kpc (red) size, and their respective
peak frequencies according to Eq.~\ref{nupeak}.  This exemplifies the
typical range of GPS and CSS sources.}
\end{figure*}

For GPS sources at lower redshifts (i.e. around unity), O'Dea (1998)
and Snellen et al. (2000) found a number of relations that relate
power, size $l$, and turn-over frequency $\nu_{\rm peak}$ of GPS
sources. The best relation exists between size and turn-over frequency
as shown in Figure \ref{odeaplot}. Converted to parameters intrinsic
to the source this can be fitted by an expression of the form

\begin{equation}\label{nupeak}
\nu_{\rm peak}=10^{(-0.21 - 0.65\times \lg{({l/{\rm kpc}})})}\;{\rm GHz},
\end{equation}
which essentially reflects some basic properties of synchrotron theory.

Similarly, one could expect that the synchrotron power is related to
the turnover frequency or the size. However, O'Dea (1998) could not
identify a clear trend. The monochromatic power at 5 GHz for currently
known GPS and CSS sources is rather constant with a large scatter. On
the other hand, Snellen et al. (2000) see a trend of decreasing flux
density with increasing size. For simplicity we here consider a
constant power at 5 GHz of $L_{\rm 5GHz}=10^{27.25}$ W/Hz in the
restframe frequency and only take redshift into account assuming a
spectrum with $\alpha=-0.6$.

It is now interesting to see how GPS sources would look like as a
function of redshift. In this respect it is important to note that
angular sizes transform differently than frequencies. Obviously, for a
fixed GPS source the turn-over frequency will monotonically decrease
with increasing redshift. On the other hand, for a modern cosmology
the angular size will decrease and later increase again at very high
$z$. This markedly different behavior is illustrated in
Fig.~\ref{sizevsz} where we place three example GPS and CSS sources at
different redshifts obeying relation \ref{nupeak}. If we take a
standard GPS source peaking at 2.7 GHz and 100 pc size, it would appear
as a compact source of 40 mas at $z\,=$ 20 -- quite common for GPS sources -- but
unusually low turn-over frequency of 125 MHz, something that is more
expected for much larger CSS sources. Of course, a source at that
redshift would also be much fainter than ``local'' GPS/CSS sources.

\begin{figure}
\centerline{\psfig{figure=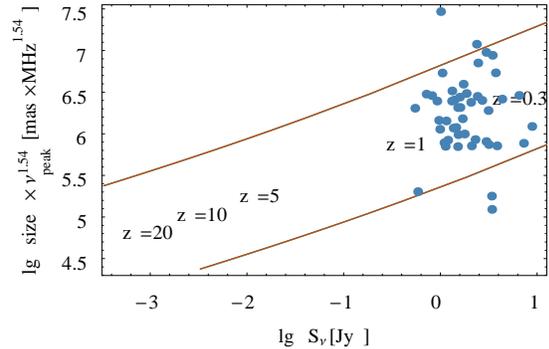,width=0.49\textwidth}}
\vspace{-1cm}
\caption{\label{sizeandnu} 
Size, frequency, and flux density roughly form a fundamental plane for
GPS radio galaxies. Here we plot a combination of frequency and size
($\nu_{\rm peak}^{1.54}\times r$) vs. the flux density for a range of
empirical GPS models and actual GPS sources from O'dea (1998). Size
and frequency for the model are coupled according to Eq.~\ref{nupeak}
for sources with different sizes of 0.1 (blue), 1 (green), and 10 kpc
(red) and fall on top of each other. The two lines delimit the scatter
in the empirical relation. The labels indicate the typical redshift of sources
in that regime. Standard GPS sources are found in the redshift range
around z=1. The bottom left corner is not occupied and is the
discovery space for young high-redshift black holes with the SKA.}
\end{figure}

We can combine the three measurables: angular size $\theta$, observed
frequency $\nu_{\rm peak,obs}$, and flux density in one diagram. One
can easily see from Eq.~\ref{nupeak} that the product of $\nu_{\rm
peak,obs}^\beta\times \theta$ should be constant for
$\beta=1/0.65=1.54$. Obviously, that parameter may have to be changed
once a better relation has been established. Using the Fanti et
al. (1990) CSS sample and the Stanghellini et al. (1996) GPS samples
with $\nu_{\rm peak}>100$ MHz we find that indeed all sources cluster
around $\left<\nu_{\rm peak,obs}^{1.54}\times
\theta\right>=10^{6.2\pm0.4}\;{\rm mas}\times{\rm MHz}^{1.54}$, 
i.e. despite their wide range in sizes and peak frequencies the sample
collapses to a cloud of sources within one order of magnitude
(Fig.~\ref{sizeandnu}).

We can then combine the predicted sizes and peak frequencies as a
function of redshift from Fig.~\ref{sizevsz}, combine them in the
same manner and overplot them on the data, extending all the way out
to high redshifts. While the current GPS/CSS sources cluster in the
top right corner, extremely high-redshift GPS/CSS are expected to
cluster in the bottom left corner -- the discovery space for the SKA.

What are the pitfalls of such a scenario? It could be that the basic
relation between size and frequency changes. This is not really
expected if the turn-over is caused by synchrotron-self absorption, as
the same physics should also apply at z=20. Modifications, however,
would be required if free-free absorption becomes important --
something that cannot be excluded in the violent early phase of galaxy
formation. On the other hand, free-free absorption in a stratified
atmosphere should lead to a qualitatively and perhaps even
quantitatively similar behavior.  Inverse Compton cooling from ambient
photons might also change the state of synchrotron sources. However,
many GPS sources are found in very luminous quasars and hence are
generally already exposed to a bright photon field. Finally, there
could be confusing sources. We know very little about the GPS/CSS
population at lower luminosities and the faint part could be heavily
confused by local low-luminosity AGN. Still, even in this case we
expect a markedly different size/frequency ratio than for sources at
very high redshifts as the size-frequency relation does not depend
strongly on power. Also, nearby galaxies could be easily excluded
through cross-correlation with existing optical catalogs.

A survey strategy for finding the first black holes would then be:
\begin{itemize}
\item a shallow all-sky multi-frequency survey in the range 100-600 MHz down to 0.1 mJy at arcsecond resolution,
\item identification of compact, highly peaked spectrum sources in that frequency range,
\item identification of empty fields in the optical,
\item re-observation to exclude variable sources,
\item observation with long baselines and resolutions of $\sim$ 10 mas to determine sizes and to pick out the ultra-compact low-frequency peaked (ULP) sources,
\item spectroscopic confirmation of remaining candidates with HI observations or by other means.
\end{itemize}

The SKA would be ideally suited for all these tasks. With a large field
of view, a broad instantaneous frequency coverage, and multibeaming it
would be very efficient in making the first surveys and generating a
candidate source list of faint, low-peaked sources. Long baselines are
then essential to confirm the AGN nature and the unusual angular sizes
expected.

\section{Summary}
In this paper we have discussed the prospects to detect and discover
black holes throughout the universe with the SKA. We have concentrated
on the compact, high-brightness temperature emission as this most
easily distinguishes black hole activity from other processes. This
radio emission is typically produced by relativistic radio jets and is
present in essentially all AGN at some level. The radio emission is
correlated with the X-ray emission, the accretion rate, and the mass of
the black hole. Supermassive black holes are best found in the radio
while stellar mass black holes are better picked up in the X-rays.
Compact jet emission seems to be more ``faithful'' for sub-Eddington
black holes and is a good tracer for faint black hole activity in the
local universe. With the SKA we can thus detect essentially all black
holes, even those as dormant as our Galactic Center, out to 40
Mpc. Moderate low-luminosity AGN can be seen out to $z=1$ and quasars
near the Eddington limit can be seen well through the Epoch of
Reionization. If black holes in the range $10^{7-9}M_\odot$ form
through accretion near the Eddington limit at redshifts $z>10$, their
cores can be picked up when they have reached a mass of
$10^7M_\odot$. This only considers the high-brightness temperature
radio cores without the extended emission and does not require beaming
which would make even fainter black holes visible.

An interesting question is whether and how the SKA can actually
identify the very first generation of black holes. Bright radio
galaxies could be used as tracers and one could search for extremely
steep spectra and/or HI absorption in the Epoch of
Reionization. However, we suggest that the first radio jets might
rather look like GPS/CSS sources with compact hot-spots as they are
stopped in the presumably dense interstellar medium of early galaxy
formation. They would reveal themselves through a low-frequency
turn-over in the spectrum due to synchrotron self-absorption and a
very compact (some ten mas) size. Since size and frequency of GPS
sources are correlated but angular size and frequency scale markedly
differently at high redshifts, the very first black holes should
occupy a rather unique place in a parameter space spanned by size,
turn-over frequency, and flux density. Discovering such sources
requires a broad bandwidth, good surveying capabilities (large
field-of-view and multi-beaming), and long baselines to measure sizes
in follow-up observations on a scale of tens of milliarcseconds. There
is, in fact, a significant chance that, due to the absorption in the
IGM in the epoch of reionization and within the assembling host
galaxy, such radio surveys with the SKA are actually the only way to
find the very first supermassive black holes in the universe.

\begin{thebibliography}{xxxxxxx}

\bibitem[\protect\citeauthoryear{{Antonucci}}{{Antonucci}}{1993}]{Antonucci199%
3}
{Antonucci} R., 1993, \araa, 31, 473

\bibitem[Becker et al.(2001)]{2001AJ....122.2850B} Becker, R.~H., et al.\ 
2001, \aj, 122, 2850

\bibitem[Blundell, Rawlings, \& Willott(1999)]{1999AJ....117..677B} 
Blundell, K.~M., Rawlings, S., \& Willott, C.~J.\ 1999, \aj, 117, 677 

\bibitem[Brunthaler et al.(2003)]{2003PASA...20..126B} Brunthaler, A., 
Falcke, H., Bower, G.~C., Aller, M.~F., Aller, H.~D., Ter{\" a}sranta, H., 
\& Krichbaum, T.~P.\ 2003, Publications of the Astronomical Society of 
Australia, 20, 126 


\bibitem[Buras, Rampp, Janka, \& Kifonidis(2003)]{2003PhRvL..90x1101B} 
Buras, R., Rampp, M., Janka, H.-T., \& Kifonidis, K.\ 2003, Physical Review 
Letters, 90, 241101 

\bibitem[Blundell, Rawlings, \& Willott(1999)]{1999AJ....117..677B} Blundell, K.~M., Rawlings, S., \& Willott, C.~J.\ 1999, \aj, 117, 677 


\bibitem[Carilli, Gnedin, \& Owen(2002)]{2002ApJ...577...22C} Carilli, C.~L., Gnedin, N.~Y., \& Owen, F.\ 2002, \apj, 577, 22 

 \bibitem[Condon(1989)]{con89}
 Condon, J.~J.\ 1989, \apj, 338, 13

\bibitem[Dennett-Thorpe, Bridle, Laing, \& 
Scheuer(1999)]{1999MNRAS.304..271D} Dennett-Thorpe, J., Bridle, A.~H., 
Laing, R.~A., \& Scheuer, P.~A.~G.\ 1999, \mnras, 304, 271 

\bibitem[\protect\citeauthoryear{{Esin}, {McClintock}, \& {Narayan}}{{Esin}
  et~al.}{1997}]{EsinMcClintockNarayan1997}
{Esin} A.~A., {McClintock} J.~E.,  {Narayan} R., 1997, \apj, 489, 865

\bibitem[\protect\citeauthoryear{{Falcke} \& {Biermann}}{{Falcke} \&
  {Biermann}}{1995}]{FalckeBiermann1995}
{Falcke} H.,  {Biermann} P.~L., 1995, \aap, 293, 665

\bibitem[Falcke, K{\" o}rding, \& Markoff(2004)]{2004A&A...414..895F} 
Falcke, H., K{\" o}rding, E., \& Markoff, S.\ 2004, \aap, 414, 895 

\bibitem[Falcke, Malkan, \& Biermann(1995)]{1995A&A...298..375F} Falcke, 
H., Malkan, M.~A., \& Biermann, P.~L.\ 1995, \aap, 298, 375 

\bibitem[Falcke \& Markoff(2000)]{2000A&A...362..113F} Falcke, H.~\& 
Markoff, S.\ 2000, \aap, 362, 113 

 \bibitem[Falcke et al.(2000)]{falet00}
 Falcke, H., Nagar, N. M., Wilson, A. S., \& Ulvestad, J. S. 2000, \apj,
 542, 197 

\bibitem[Fan et al. (2003)]{2003AJ....125.1649F} Fan, X., et al.\ 2003, \aj, 
125, 1649 

\bibitem[]{}Fanti, R., Fanti, C., Schilizzi, R. T., Spencer, R. E., Rendong, N., Parma, P., van Breugel, W. J. M., \& Venturi, T. 1990b, A\&A, 231, 333

\bibitem[\protect\citeauthoryear{{Fender}}{{Fender}}{2001}]{Fender2001}
{Fender} R.~P., 2001, \mnras, 322, 31

\bibitem[]{FenderBelloni2004}{Fender} R.~P., \& Belloni, T., 2004, ARA\&A, 42, 317-364

\bibitem[\protect\citeauthoryear{{Fender}, {Gallo}, \& {Jonker}}{{Fender}
  et~al.}{2003}]{FenderGalloJonker2003}
{Fender} R.~P., {Gallo} E.,  {Jonker} P.~G., 2003, \mnras, 343, L99

\bibitem[Ferrarese \& Merritt(2000)]{2000ApJ...539L...9F} Ferrarese, L.~\& 
Merritt, D.\ 2000, \apjl, 539, L9 

\bibitem[\protect\citeauthoryear{{Fossati} et~al.}{{Fossati}
  et~al.}{1998}]{FossatiMaraschiCelotti1998}
{Fossati} G., {Maraschi} L., {Celotti} A., {Comastri} A.,  {Ghisellini} G.,
  1998, \mnras, 299, 433

\bibitem[\protect\citeauthoryear{{Gallo}, {Fender}, \& {Pooley}}{{Gallo}
  et~al.}{2003}]{GalloFenderPooley2003}
{Gallo} E., {Fender} R.~P.,  {Pooley} G.~G., 2003, \mnras, 344, 60

\bibitem[Gebhardt et al.(2000)]{2000ApJ...539L..13G} Gebhardt, K., et al.\ 
2000, \apjl, 539, L13 

\bibitem[\protect\citeauthoryear{{Ghisellini} \& {Celotti}}{{Ghisellini} \&
  {Celotti}}{2001}]{GhiselliniCelotti2001}
{Ghisellini} G.,  {Celotti} A., 2001, \aap, 379, L1

\bibitem[\protect\citeauthoryear{{Gursky} \& {Schwartz}}{{Gursky} \&
  {Schwartz}}{1977}]{GurskySchwartz1977}
{Gursky} H.,  {Schwartz} D.~A., 1977, \araa, 15, 541

 \bibitem[Ho et al.(1997)]{hoet97}
 Ho, L. C., Filippenko, A. V., \& Sargent, W. L. W. 1997a,
 \apjs, 112, 315 

 \bibitem[Ho, Filippenko, \& Sargent(2003)]{hoet03}
 Ho, L.~C., Filippenko, A.~V., \& Sargent, W.~L.~W.\ 2003, \apj, 583, 159

\bibitem[Kogut et al.(2003)]{2003ApJS..148..161K} Kogut, A., et al.\ 2003, 
\apjs, 148, 161 

 \bibitem[Kukula et al.(1995)]{kuket95}
 Kukula, M. J., Pedlar, A., Baum, S. A., \& O'Dea, C. P.
 1995, \mnras, 276, 1262

\bibitem[Maccarone(2004)]{2004MNRAS.351.1049M} Maccarone, T.~J.\ 2004, 
\mnras, 351, 1049 

\bibitem[\protect\citeauthoryear{{Maccarone}}{{Maccarone}}{2003}]{Maccarone200%
3}
{Maccarone} T.~J., 2003, \aap, 409, 697

\bibitem[\protect\citeauthoryear{{Maccarone} \& {Coppi}}{{Maccarone} \&
  {Coppi}}{2003}]{MaccaroneCoppi2003}
{Maccarone} T.~J.,  {Coppi} P.~S., 2003, \mnras, 338, 189

\bibitem[\protect\citeauthoryear{{Markoff}, {Falcke}, \& {Fender}}{{Markoff}
  et~al.}{2001}]{MarkoffFalckeFender2001}
{Markoff} S., {Falcke} H.,  {Fender} R., 2001, \aap, 372, L25

\bibitem[\protect\citeauthoryear{{Markoff} et~al.}{{Markoff}
  et~al.}{2003}]{MarkoffNowakCorbel2003}
{Markoff} S., {Nowak} M., {Corbel} S., {Fender} R.,  {Falcke} H., 2003, \aap,
  397, 645

\bibitem[\protect\citeauthoryear{{Melia} \& {Falcke}}{{Melia} \&
  {Falcke}}{2001}]{MeliaFalcke2001}
{Melia} F.,  {Falcke} H., 2001, \araa, 39, 309

\bibitem[\protect\citeauthoryear{{Merloni}, {Heinz}, \& {di Matteo}}{{Merloni}
  et~al.}{2003}]{MerloniHeinzdiMatteo2003}
{Merloni} A., {Heinz} S.,  {di Matteo} T., 2003, \mnras, 345, 1057

 \bibitem[Meurs \& Wilson(1984)]{meuwil84}
  Meurs, E.~J.~A.~\& Wilson, A.~S.\ 1984, \aap, 136, 206

\bibitem[\protect\citeauthoryear{{Mirabel} \& {Rodr{\' i}guez}}{{Mirabel} \&
  {Rodr{\' i}guez}}{1999}]{MirabelRodriguez1999}
{Mirabel} I.~F.,  {Rodr{\' i}guez} L.~F., 1999, \araa, 37, 409

 \bibitem[Nagar et al.(2000)]{naget00}
 Nagar, N. M., Falcke, H., Wilson, A. S., \& Ho, L. C.
 2000, \apj, 542, 186 

 \bibitem[Nagar et al.(2002)]{naget02}
 Nagar, N.~M., Falcke, H., Wilson, A.~S., \& Ulvestad, J.~S.\ 2002, \aap, 392, 53
 
 \bibitem[Nagar et al.(2004)]{naget04}
 Nagar, N.~M., et al, 2004, \aap, submitted

\bibitem[\protect\citeauthoryear{{Narayan} \& {Yi}}{{Narayan} \&
  {Yi}}{1995}]{NarayanYi1995}
{Narayan} R.,  {Yi} I., 1995, \apj, 452, 710

\bibitem[\protect\citeauthoryear{{Nowak}}{{Nowak}}{1995}]{Nowak1995}
{Nowak} M.~A., 1995, \pasp, 107, 1207

\bibitem[\protect\citeauthoryear{{O'Dea}}{{O'Dea}}{1998}]{O'Dea1998}
{O'Dea} C.~P., 1998, \pasp, 110, 493

\bibitem[]{}O'Dea, C. P., \& Baum, S. A. 1997, AJ, 113, 148

\bibitem[Portegies Zwart et al.(2004)]{2004Natur.428..724P} Portegies Zwart, S.~F., 
Baumgardt, H., Hut, P., Makino, J., \& McMillan, S.~L.~W.\ 2004, \nat, 428, 
724 

\bibitem[\protect\citeauthoryear{{Poutanen}}{{Poutanen}}{1998}]{Poutanen1998}
{Poutanen} J., 1998, in Theory of Black Hole Accretion Disks, Cambridge
  University Press, p. 100

\bibitem[\protect\citeauthoryear{{Rawlings} \& {Saunders}}{{Rawlings} \&
  {Saunders}}{1991}]{RawlingsSaunders1991}
{Rawlings} S.,  {Saunders} R., 1991, \nat, 349, 138

 \bibitem[Sandage, Tammann, \& Yahil(1979)]{sanet79}
 Sandage, A., Tammann, G.~A., \& Yahil, A.\ 1979, \apj, 232, 352

\bibitem[\protect\citeauthoryear{{Sanders}}{{Sanders}}{1983}]{Sanders1983}
{Sanders} R.~H., 1983, \apj, 266, 73

\bibitem[\protect\citeauthoryear{{Shakura} \& {Sunyaev}}{{Shakura} \&
  {Sunyaev}}{1973}]{ShakuraSunyaev1973}
{Shakura} N.~I.,  {Sunyaev} R.~A., 1973, \aap, 24, 337

\bibitem[\protect\citeauthoryear{{Shapiro}, {Lightman}, \& {Eardley}}{{Shapiro}
  et~al.}{1976}]{ShapiroLightmanEardley1976}
{Shapiro} S.~L., {Lightman} A.~P.,  {Eardley} D.~M., 1976, \apj, 204, 187

\bibitem[Silk \& Rees(1998)]{1998A&A...331L...1S} Silk, J.~\& Rees, M.~J.\ 
1998, \aap, 331, L1 

\bibitem[Snellen et al.(2000)]{2000MNRAS.319..445S} Snellen, I.~A.~G., 
Schilizzi, R.~T., Miley, G.~K., de Bruyn, A.~G., Bremer, M.~N., \& R{\"o}ttgering, H.~J.~A.\ 2000, \mnras, 319, 445 

\bibitem[]{}Stanghellini, C., Dallacasa, D., O'Dea, C. P., Baum, S. A., Fanti, R., \& Fanti, C. 1996, in: Proc. Second Workshop on Gigahertz Peaked-Spectrum and Compact Steep-Spectrum Radio Sources, ed. I. A. G. Snellen, R. T. Schilizzi, H. J. A. Röttgering, \& M. N. Bremer (Leiden: Leiden Obs.), 4

\bibitem[\protect\citeauthoryear{{Sunyaev} \& {Tr\"umper}}{{Sunyaev} \&
  {Tr\"umper}}{1979}]{SunyaevTruemper1979}
{Sunyaev} R.~A.,  {Tr\"umper} J., 1979, \nat, 279, 506

 \bibitem[Ulvestad \& Ho(2001)]{ulvho01a}
 Ulvestad, J.~S.~\& Ho, L.~C.\ 2001, \apj, 558, 561

\bibitem[\protect\citeauthoryear{{Urry} \& {Padovani}}{{Urry} \&
  {Padovani}}{1995}]{UrryPadovani1995}
{Urry} C.~M.,  {Padovani} P., 1995, \pasp, 107, 803

\bibitem[\protect\citeauthoryear{{von Montigny} et~al.}{{von Montigny}
  et~al.}{1995}]{MontignyBertschChiang1995}
{von Montigny} C., {Bertsch} D.~L., {Chiang} J., et~al., 1995, \apj, 440, 525


\bibitem[Willott, McLure, \& Jarvis(2003)]{2003ApJ...587L..15W} Willott, 
C.~J., McLure, R.~J., \& Jarvis, M.~J.\ 2003, \apjl, 587, L15 


\bibitem[\protect\citeauthoryear{{Yuan}, {Markoff}, \& {Falcke}}{{Yuan}
  et~al.}{2002}]{YuanMarkoffFalcke2002}
{Yuan} F., {Markoff} S.,  {Falcke} H., 2002, \aap, 383, 854

\end{thebibliography}

\end{document}